\documentclass[aps,pra,twocolumn,showpacs,preprintnumbers,amsmath,amssymb]{revtex4-2}
\usepackage{soul}
\usepackage{amsmath}

\usepackage{natbib}
\usepackage{graphicx}
\usepackage{color}
\usepackage{tabularx}
\usepackage{multirow}

\usepackage{wasysym}

\usepackage{amssymb}

\usepackage[utf8]{inputenc}

\usepackage{float}

\usepackage{subfigure}

\usepackage{float}

\usepackage{hyperref}

\usepackage{cancel}

\usepackage{xcolor}

\usepackage{setspace}
\usepackage{soul}

\begin{document}

\title{Effect of depolarizing and quenching collisions\\
on contrast of the coherent population trapping resonance}

\author{K.\,M.~Sabakar$^{1}$}
\author{M.\,I.~Vaskovskaya$^{1}$}
\author{D.\,S.~Chuchelov$^{1}$}
\author{E.\,A.~Tsygankov$^{1}$}
\email[]{tsygankov.e.a@yandex.ru}
\author{V.\,V.~Vassiliev$^{1}$}
\author{S.\,A.~Zibrov$^{1}$}
\author{V.\,L.~Velichansky$^{1,2}$}
\affiliation{1. Lebedev Physical Institute of the Russian Academy of Sciences, Leninsky Prospect 53, Moscow, 119991 Russia}
\affiliation{2.~National~Research~Nuclear~University~MEPhI,~Kashirskoye~Highway~31,~Moscow,~115409~Russia}

\begin{abstract}
We investigate the effect of buffer gases on the coherent population trapping resonance induced by a $\sigma$-polarized optical field in $^{87}$Rb atoms. Our experimental results show that inert gases, which depolarize the excited state of the alkali-metal atoms, provide higher contrast than nitrogen that effectively quenches their fluorescence. We also demonstrate that elimination of the spontaneous radiation does not significantly decrease the width at moderate temperatures of an atomic medium. Therefore, a mixture of inert gases can be preferable over a mixture with nitrogen for atomic clocks.
\end{abstract}

\maketitle

All-optical interrogation schemes utilizing the effect of coherent population trapping (CPT)~\cite{ARIMONDO1996257}, progress in miniature vapor-cell production technology and advances in diode lasers~\cite{michalzik2013vcsel} have led to the development of chip-scale atomic clocks~\cite{10.1117/12.726792}.
Their main advantages over other frequency standards are smaller size and lower power consumption, but they have lower frequency stability. Currently, many research groups are seeking for new approaches to improve the long-term frequency stability of such atomic clocks~\cite{Zhang:s,8786876,yanagimachi,gozzelino2020kr}. In the absence of a frequency drift, the stability is proportional to $1/\sqrt{\tau}$, where $\tau$ is the averaging time~\cite{shah2010advances}. In this case, a further improvement of the long-term frequency stability can be achieved only by the increase of the short-term stability, which depends on the contrast-to-width-ratio of the CPT resonance. In what follows, we call it the quality factor, or Q-factor.

The standard approach to reduce the relaxation rate of the ground-state coherence 
occurring due to collisions of alkali-metal atoms with atomic cell walls is the usage of a buffer gas. It provides diffusion of atoms to the walls with a slower speed than their unperturbed movement with the thermal velocity. The probability to lose the polarization upon collisions with buffer gas particles is less than under interaction with the cell walls. However, an increase of the buffer gas pressure results eventually in the collisional rebroadening of the CPT resonance. These opposite dependencies give the value of buffer gas pressure that provides the minimum width. This value depends on the dimensions and geometry of the cell~\cite{vanier1989quantum}.
The buffer gas induces a temperature-dependent shift of the CPT resonance frequency, which can be suppressed by using a mixture of two gases with linear temperature coefficients of opposite signs~\cite{doi:10.1063/1.331467}.
Most often, a mixture of argon and nitrogen is used.

\begin{figure}[t]
\centering 
\includegraphics[width=\columnwidth]{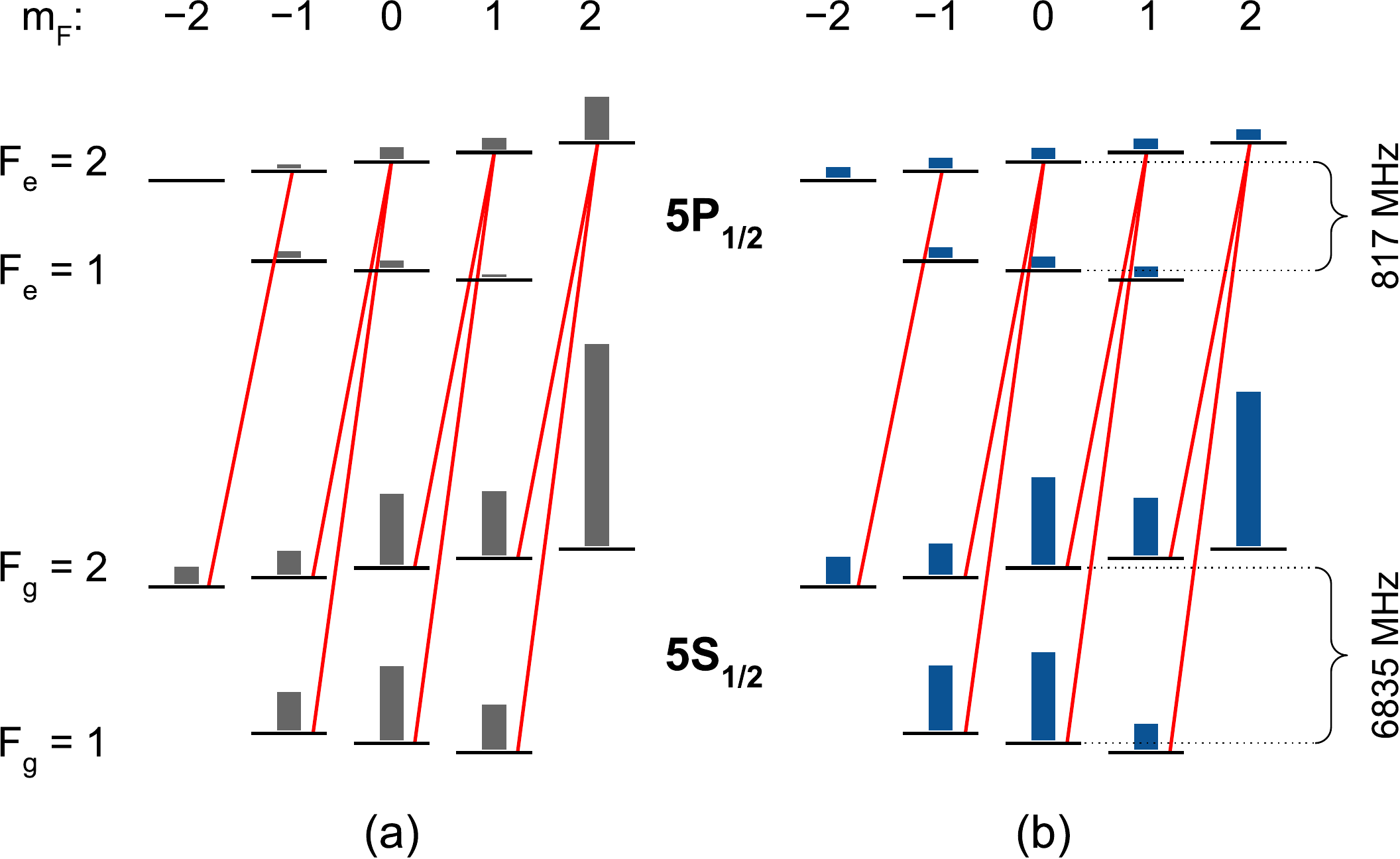}
\caption{Energy level structure and electric-dipole transitions to sublevels of hyperfine component \hbox{$F_e=2$} induced by an optical field with $\sigma^+$ polarization. Columns show distribution of populations over $5$S$_{1/2}$ and $5$P$_{1/2}$ states. They were calculated in a model without (a) and with (b) accounting for the depolarization; see \hyperref[Appendix]{Appendix} for details. The heights of the columns for the excited state are increased by about five orders of magnitude.}
\label{LevelsScheme}
\end{figure}

\vbox{It is known that inert gases depolarize the excited state of alkali-metal atoms and tend to equalize populations of its magnetic sublevels~\cite{franzen1957atomic,zhitnikov1970optical,okunevich1970relaxation,RevModPhys.44.169}. The equalization of populations should increase the CPT resonance amplitude detected in $\sigma^+$-$\sigma^+$ scheme.
Indeed, the depolarization reduces the number of atoms pumped to the sublevel $5$P$_{1/2}$ \hbox{$m_{F_e}=2$} (we consider the case of $^{87}$Rb atoms). 
Therefore, a smaller amount of atoms is optically pumped to the non-absorbing sublevel \hbox{$m_{F_g}=2$} of the ground state and more atoms arrive at the working sublevels \hbox{$F_g={1,\,2},\,m_{F_g}=0$} due to spontaneous transitions.
Fig.~\ref{LevelsScheme} demonstrates distribution of populations over magnetic sublevels for the case of zero and complete excited-state depolarization.
They were obtained by solving density-matrix equations accounting for all electric-dipole transitions of D$_1$ line induced by a bichromatic $\sigma^+$-polarized optical field. 
Power broadening of the CPT resonance was set to be $3$-times greater than relaxation rate of ground-state elements to make difference in populations evident. Details of calculations are given in \hyperref[Appendix]{Appendix}.}

Nitrogen quenches fluorescence of alkali-metal atoms due to the transfer of the excited-state energy to molecular vibrations. This prevents broadening of the CPT resonance induced by the spontaneous radiation, therefore, nitrogen is often considered as a preferable buffer gas for atomic clocks~\cite{vanier1989quantum,doi:10.1063/1.5026238}. Transitions from the excited to the ground states during quenching have the same selection rules as spontaneous decay~\cite{PhysRevA.11.1,PhysRevA.25.2985}, but the effect of nitrogen on the population distribution of the excited state has not been studied in detail and is poorly described in the literature.

We assume that molecular gases can 
to some extent prevent the excited-state depolarization. In this case, nitrogen should reduce the CPT resonance amplitude compared to inert gases while improving its width due to the elimination of the spontaneous radiation. 
The influence of these two factors on Q-factor is opposite. The goal of this paper is to estimate which of them is more significant. To check this, we compared the contrast and width of the CPT resonance in \hbox{Ar, Ne, and N$_2$}.

\section{Experiment}
\label{SectionExamples}

\begin{figure}[b]
\centering 
\includegraphics[width=\columnwidth]{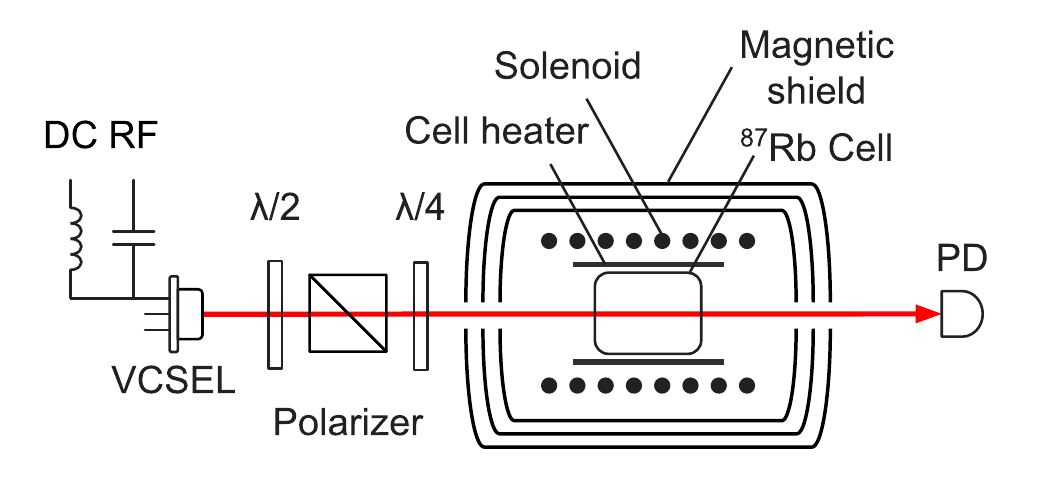}
\caption{The layout of experimental setup.}
\label{ExpSetup}
\end{figure}

The experimental setup is schematically shown in Fig.~\ref{ExpSetup}. We used a single-mode vertical-cavity surface-emitting laser generating at \hbox{$\simeq795$~nm}. The DC and RF components of the injection current were fed to the laser via a bias tee. The modulation frequency was close to $3.417$~GHz, and the first sidebands of the polychromatic optical field were tuned to transitions $F_g=2\rightarrow F_e=2$, $F_g=1\rightarrow F_e=2$ of the $^{87}$Rb D$_1$~line. The power of the RF field was set to provide the highest amplitudes of the first-order sidebands. A polarizer and a quarter-wave plate were used to form the CPT resonance in the $\sigma^+$-$\sigma^+$ scheme. The diameter of the laser beam was $3$~mm. The laser wavelength was stabilized by a feedback loop that controls the temperature of the laser diode.

An atomic cell was placed in a longitudinal magnetic field of $0.02$~G to separate the metrological CPT resonance from magneto-sensitive ones at the transitions between sublevels $m_{F_g}=\pm1$. The temperature of the atomic cell was maintained with an accuracy of $0.01$~$^{\circ}$C. The cell, heater, and solenoid were placed in a three-layer $\mu$-metal magnetic shield, providing better than $500$-fold suppression of the laboratory magnetic field.

We have manufactured three sets of cylindrical atomic cells with CO$_2$ laser-welded windows ($8$~mm diameter, $15$~mm length, $0.7$~mm wall thickness) and filled them with isotopically enriched $^{87}$Rb and one of the buffer gases: N$_2$, Ar, or Ne. The buffer gas pressures are $30$, $60$, and $90$~Torr. We used pinch-off glass welding to seal the stem at a distance of about $20$~mm from the cell body so as not to heat it. This ensures that the actual gas pressure inside the cell differs from the pressure in the filling chamber by no more than $1$\%.

\begin{figure}[b]
\centering 
\includegraphics[width=\columnwidth]{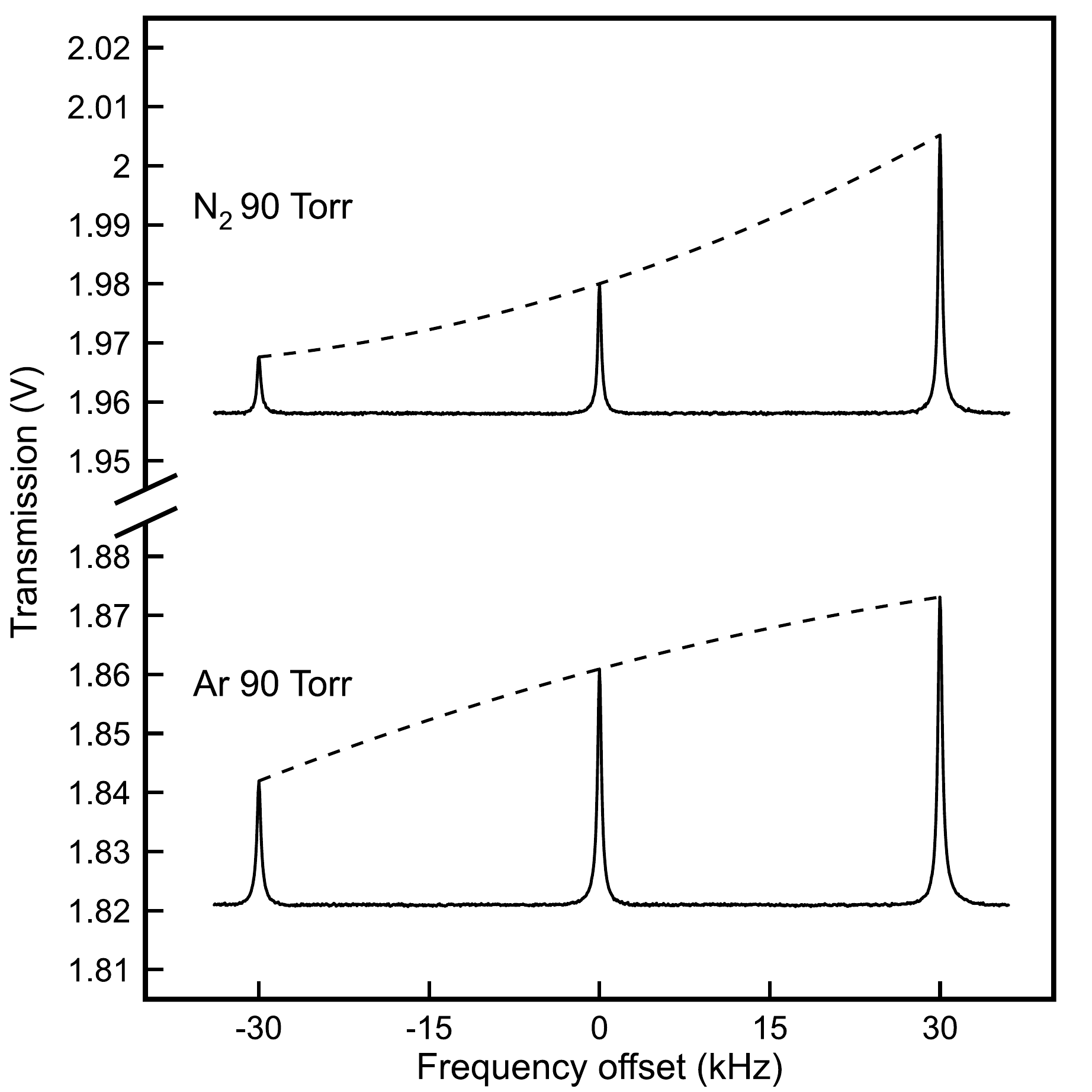}
\caption{Metrological (central) and magneto-sensitive CPT resonances obtained in atomic cells filled with nitrogen and argon at a pressure of $90$~Torr. The dashed lines serve as a guide for the eye and show the difference in the amplitudes of the resonances.}
\label{Magneto-sensitive}
\end{figure}

\begin{figure*}[t]
\centering 
\includegraphics[width=\textwidth]{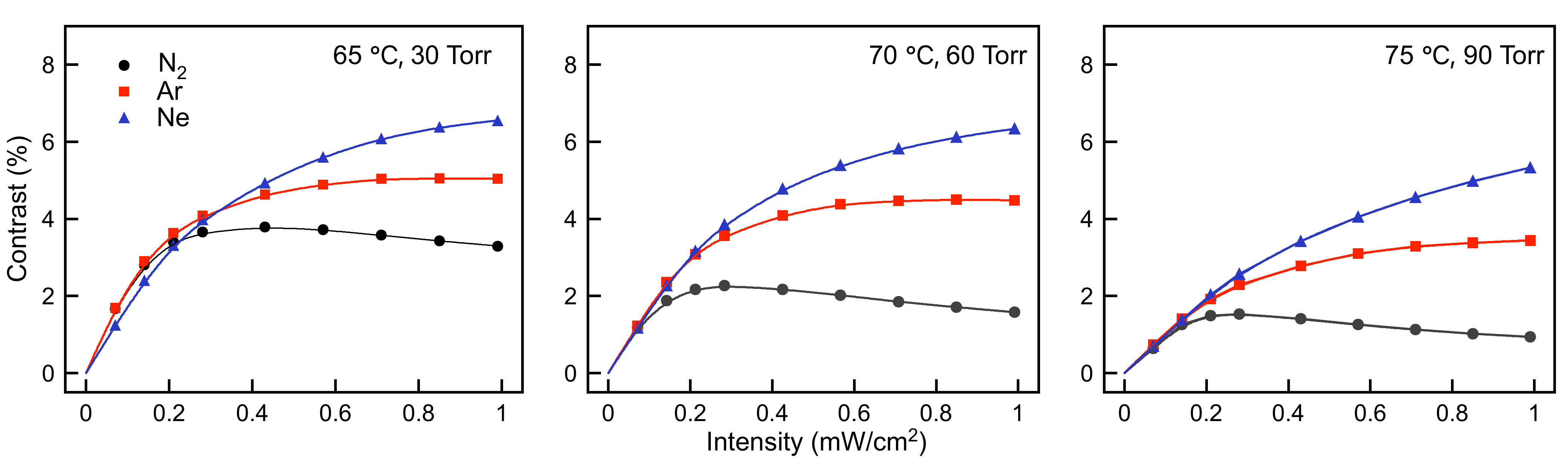}
\caption{Dependencies of the CPT resonance contrast on the laser intensity for different temperatures and pressures of nitrogen, argon, and neon.}
\label{Contrasts}
\end{figure*}

\begin{figure*}[t]
\centering 
\includegraphics[width=\textwidth]{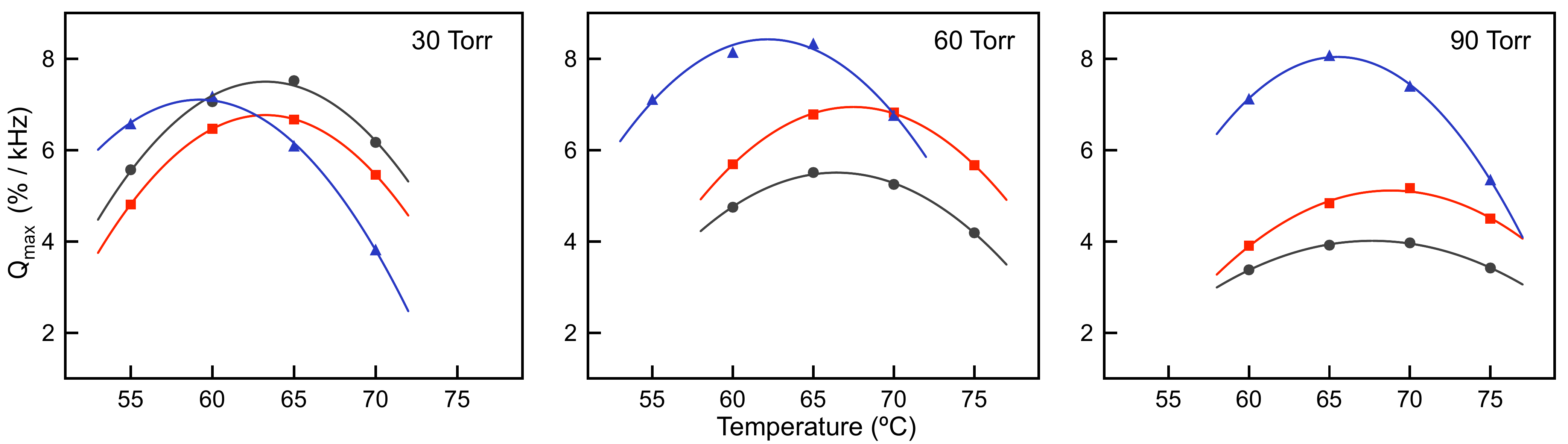}
\caption{Dependencies of the maximal (in terms of intensity) quality factor Q$_{\text{max}}$ of the CPT resonance on the cell temperature for different pressures of nitrogen, argon, and neon. The legend is the same as in Fig.~\ref{Contrasts}.}
\label{Qfactors}
\end{figure*}

Fig.~\ref{Magneto-sensitive} shows metrological and magneto-sensitive CPT resonances obtained in two atomic cells filled with nitrogen and argon at a pressure of $90$~Torr. The inhomogeneity of the magnetic field did not lead to a noticeable broadening of magneto-sensitive resonances, thus, we consider their amplitudes to be determined by populations of the corresponding magnetic sublevels. Experimental conditions are the same for both cells: temperature and optical field intensity are $65$~$^\circ$C and $0.3$~mW/cm$^2$. The non-resonant radiation losses in both cells are almost equal. However, some differences in the signals can be seen. First, resonances on sublevels $F_g=1,2, m_{F_g}=-1$ (left) and $F_g=1,2, m_{F_g}=0$ (central) have noticeably greater amplitudes in Ar than in N$_2$. Second, the background transmission in nitrogen is higher when the microwave frequency is detuned from the CPT resonance ($\simeq1.96$~V in N$_2$, $\simeq1.82$~V in Ar). On the contrary, nitrogen should provide a slightly smaller transmission level due to the lower collisional broadening of the $^{87}$Rb D$_1$~line~\cite{pitz2014pressure}. We attribute the mentioned above features to the negative impact of the fluorescence quenching in nitrogen, which reduces efficiency of the excited state depolarization and enhances pumping to the non-absorbing sublevel.

Dependencies of the contrast (the ratio of the resonance amplitude to the transmission level at resonance peak) of the metrological CPT resonance on the laser field intensity for all pressures are shown in Fig.~\ref{Contrasts}. The difference in contrast between gases is negligible at intensities below~$0.1$~mW/cm$^{2}$ for all pressures. As the intensity increases, the rate of optical pumping becomes significant and the divergence between dependencies arises. For N$_2$, the contrast reaches a maximum and then slightly decreases. For Ar and Ne the contrast does not decrease even at the highest available intensity of $1$~mW/cm$^{2}$. Neon provides the highest contrast, reaching a value of $6.5$\%, while the maximal contrasts for argon and nitrogen are $5$\% and $3.6$\%, respectively. s the pressure increases, the dependencies remain almost the same and the relation $C_{\text{max}}^{\text{Ne}}>C_{\text{max}}^{\text{Ar}}>C_{\text{max}}^{\text{N}_2}$ does not change, but the maximal contrasts decline. It happens due to the growth of homogeneous broadening that leads to decrease in amount of atoms optically pumped into the dark state. The increase in temperature cannot fully compensate the loss in number of atoms since the relaxation rate of the ground-state coherence becomes greater due to the spin-exchange mechanism. Therefore, the maximal absorption contrast is achieved at higher temperature for greater buffer gas pressure and falls with growth of the latter. The difference in contrasts between the inert gases is due to the smaller broadening of $^{87}$Rb D$_1$ line by Ne~\cite{pitz2014pressure}.
\begin{figure*}[t]
\centering 
\includegraphics[width=\textwidth]{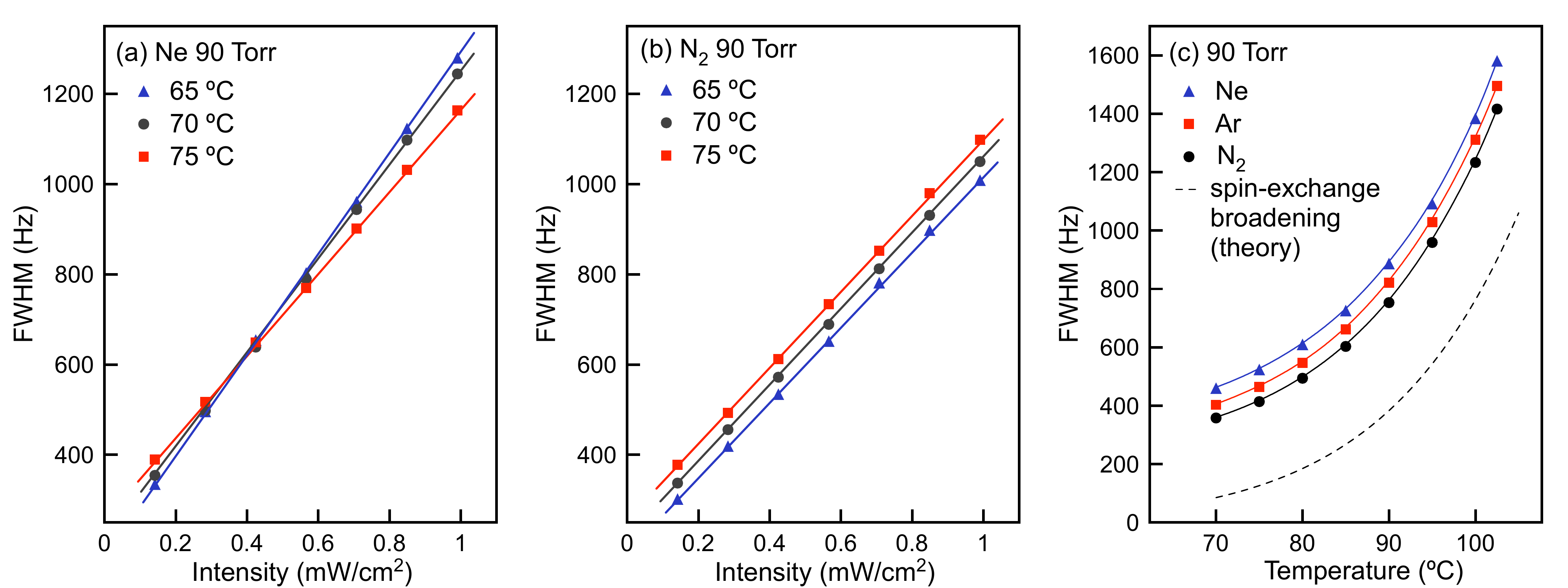}
\caption{Dependencies of the CPT resonance width on the optical field intensity for different temperatures in neon~(a) and nitrogen~(b) and on the temperature for atomic cells with nitrogen, argon and neon~(c). Pressure of all buffer gases is $90$~Torr. The dashed line in (c) is the spin-exchange broadening for $^{87}$Rb calculated via~Eq.~\eqref{SE}.}
\label{IWidth}
\end{figure*}

Similarly to the contrast, the dependencies of the CPT resonance width and Q-factor on the intensity were obtained. From each dependence Q(I) we defined the maximum value Q$_{\text{max}}$ and plotted the dependence of the Q$_{\text{max}}$ on temperature for each gas and all pressures (Fig.~\ref{Qfactors}). Under the same conditions, the resonance in nitrogen has the smallest width. Nitrogen has an advantage over argon and neon of about $20$\% and $10$\% in Q$_{\text{max}}$ for a pressure of $30$~Torr, which is achieved due to the narrower resonance. At higher buffer gas pressures, the advantage of inert gases in contrast exceeds the advantage of nitrogen in width. As a result, Q$^{\text{Ne}}_{\text{max}}$/Q$^{\text{N}_{2}}_{\text{max}}$ is close to $1.6$ at $60$~Torr and to $2$ at $90$~Torr. Note that neon maximizes Q-factor at lower temperatures than other gases at all pressures.

Fig.~\ref{IWidth}~(a,$\,$b) shows dependencies of the CPT resonance width on the optical field intensity for different temperatures in neon and nitrogen. In neon the width decreases with temperature at intensities above $0.4$~mW/cm$^2$. This feature is due to the light narrowing effect~\cite{godone2002dark}, which takes place when a sufficiently large number of atoms are coherently trapped in a dark state. In nitrogen, as the temperature increases, the width increases in the same way for all intensities, which indicates the much smaller impact of this effect. Therefore, more atoms resides at the non-absorbing sublevel in N$_2$ compared to Ne.

We have studied the potential benefit of fluorescence quenching in nitrogen for the CPT resonance width. For this, we made three additional cells with the same diameter of $8$~mm, but smaller internal length of $2.5$~mm. The decrease of length prevents complete absorption of low-intensity laser radiation at high temperatures and Rb concentrations, when the influence of spontaneous photons on the width should be more evident. The cells were filled with $90$~Torr of Ar, Ne, and N$_2$. The beam was expanded to $6$~mm in diameter and the operating radiation intensity was $0.1$~mW/cm$^2$. The measured dependencies of the CPT resonance width on temperature in these cells are shown in Fig.~\ref{IWidth}(c). As one would expect, the width should grow faster with temperature in inert gases than in nitrogen due to the spontaneous radiation. However, dependencies in all gases reveal the same behavior, which is typical for spin-exchange broadening. The contribution of this mechanism to the coherence relaxation is given by~\cite{RevModPhys.44.169}
\begin{equation}
\Gamma_{se}=\dfrac{6I+1}{8I+4}\sigma_{se}v_rn,
\label{SE}
\end{equation}

\noindent where $I$ is the nuclear spin, $\sigma_{se}$ is the spin-exchange cross-section, $v_r$ is the average relative velocity, and $n$ is the concentration of the alkali-metal atoms. The dashed line in Fig.~\ref{IWidth}(c) is the dependence of $\Gamma_{se}/\pi$ on temperature for $^{87}$Rb plotted for concentration taken from~\cite{Steck} and the most reliable value of $\sigma_{se}$ equal to $1.9\cdot10^{-14}$~cm$^2$~\cite{walter2002magnetic}. However, we have not observed a noticeable broadening of the CPT resonance in inert gases at high temperatures compared to nitrogen. Similar result was obtained earlier in~\cite{knappe2002temperature} for $^{133}$Cs in the temperature range $20$--$65$~$^{\circ}$C. Thus, we do not associate the difference in widths with the quenching effect, it is probably related to the lower diffusion coefficient in nitrogen~\cite{Arditi&Carver,Pouliot}, which determines the rate of Rb coherence relaxation as a result of collisions with the cell walls.

\section{Discussion}

Considering the choice of a buffer gas for CPT-based atomic clocks, we believe that a proper mixture of Ar and Ne is preferable to a mixture containing N$_2$. When using inert gases, it is possible to achieve a higher resonance contrast due to depolarization of $^{87}$Rb excited state. Moreover, the maximal Q-factor of the resonance in neon is achieved at a lower temperature than in nitrogen, which reduces the clock power consumption.

Another advantage of Ar-Ne mixture is the ability to suppress the light shift at higher buffer gas pressures. As we demonstrated in~\cite{BGpaper}, the AC Stark shift of the CPT resonance frequency cannot be eliminated if the homogeneous broadening of optical lines exceeds a certain value. Since $^{87}$Rb D$_1$ line collisional broadening rate for Ne is about $1.5$~times smaller than that for N$_2$ ~\cite{pitz2014pressure}, it is possible to obtain the minimal relaxation rate of the coherence and retain the ability to suppress the light shift in miniature atomic cells, which is significant for chip-scale atomic clocks.

Finally, the depolarization of the $^{87}$Rb excited state in inert gases leads to a smaller difference in populations of the ground-state working sublevels due to the repopulation pumping mechanism. Unequal populations are the source of the CPT resonance asymmetry and the nonlinear dependence of its frequency on the laser field intensity~\cite{https://doi.org/10.48550/arxiv.2012.13731}. This hinders the light shift suppression methods based on the laser field intensity modulation (see, for example,~\cite{doi:10.1063/1.2360921}).

\section{Summary}

We have demonstrated that argon and neon provide a higher contrast of the CPT resonance than nitrogen in the $\sigma^+$-$\sigma^+$ scheme. The difference in contrast is significant when the optical pumping rate dominates over the ground-state relaxation. We explain this effect as follows. Quenching of alkali-metal atoms fluorescence reduces the degree of the excited-state depolarization, which increases the population of the non-absorbing sublevel. As a result, the amount of atoms that can be optically pumped to the dark state becomes smaller and the amplitude of CPT resonance decreases. We have not found a benefit from quenching of the fluorescence for the width of the CPT resonance at temperatures providing the maximal Q-factor. The difference in Q-factor between Ne and N$_2$ increases with the buffer gas pressure, reaching a factor of $2$ at $90$~Torr. Hence, a mixture of inert gases can be more advantageous for CPT-based atomic clocks than a mixture with nitrogen.

\section{Acknowledgments}
\label{Acknowledgments}

The authors receive funding from Russian Science Foundation (grant No.~19-12-00417).

\section{Appendix}
\label{Appendix}
Here we consider the following model of optical pumping in four-level system with moments $F=2,1$ in excited and ground states; see Fig.~\ref{LevelsScheme}. Components of bichromatic optical field
\begin{equation*}
\mathbf{E}(t)=\mathbf{e}\dfrac{\mathcal{E}}{2}\left(e^{-i\omega_rt}+e^{-i\omega_bt}+c.c.\right),
\end{equation*}

\noindent where $\mathbf{e}=(\mathbf{e}_x+i\mathbf{e}_y)/\sqrt{2}$, induce transitions between levels $F_g=2$, $F_e=2$ and $F_g=1$, $F_e=2$, respectively, having optical detuning $\Delta$. The corresponding detuning from $F_e=1$ level is $\Delta+\omega_e$, where $\omega_e$ is the hyperfine splitting of the excited state. Frequency spacing between components is close to hyperfine splitting of the ground state: $\omega_b-\omega_r=\omega_g+\delta$, where detuning $\delta$ is much smaller than $\omega_g$. Phenomenological relaxation constant $\Gamma$ was introduced in equations for optical coherences to account for homogeneous broadening of absorption line occurring under collision of alkali-metal atoms with particles of a buffer gas. We assume that $\gamma\ll\Gamma$, where $\gamma$ denotes natural width of the excited state. Rabi frequency $V=d\mathcal{E}/2\hbar$ contains the reduced dipole matrix element. For simplicity, one phenomenological constant $\Gamma_g$ is used to describe relaxation of the ground-state sublevels. Finally, we do not account for magneto-sensitive CPT resonances and consider only the one between sublevels $m_{F_g}=0$. Initial equations for elements of the density matrix are solved under approximations of low saturation regime and the resonant one for optical field (also known as the rotating-wave approximation), which allows us to obtain the following system for the steady-state regime.

Namely, there are equations for populations of the excited state under absence of its depolarization:
\begin{subequations}
\begin{equation}
\rho^{uu}_{-2-2}=0,
\end{equation}
\begin{equation}
\rho^{uu}_{-1-1}=\dfrac{V^2}{6}\dfrac{\Gamma/(\gamma/2)}{\Delta^2+\Gamma^2}\rho^{22}_{-2-2},
\end{equation}
\begin{equation}
\rho^{dd}_{-1-1}=\dfrac{V^2}{2}\dfrac{\Gamma/(\gamma/2)}{(\Delta+\omega_e)^2+\Gamma^2}\rho^{22}_{-2-2},
\end{equation}
\begin{equation}
\rho^{uu}_{00}=\dfrac{V^2}{4}\dfrac{\Gamma/(\gamma/2)}{\Delta^2+\Gamma^2}\left(\rho^{22}_{-1-1}+\dfrac{1}{3}\rho^{11}_{-1-1}\right),
\end{equation}
\begin{equation}
\rho^{dd}_{00}=\dfrac{V^2}{4}\dfrac{\Gamma/(\gamma/2)}{(\Delta+\omega_e)^2+\Gamma^2}\left(\rho^{22}_{-1-1}+\dfrac13\rho^{11}_{-1-1}\right),
\end{equation}
\begin{equation}
\rho^{uu}_{11}=\dfrac{V^2}{4}\dfrac{\Gamma/(\gamma/2)}{\Delta^2+\Gamma^2}\left[\rho^{22}_{00}+\rho^{11}_{00}-2\text{Re}\left(\rho^{21}_{00}\right)\right],
\end{equation}
\begin{equation}
\rho^{dd}_{11}=\dfrac{V^2}{12}\dfrac{\Gamma/(\gamma/2)}{(\Delta+\omega_e)^2+\Gamma^2}\left[\rho^{22}_{00}+\rho^{11}_{00}-2\text{Re}\left(\rho^{21}_{00}\right)\right],
\end{equation}
\begin{equation}
\rho^{uu}_{22}=\dfrac{V^2}{2}\dfrac{\Gamma/(\gamma/2)}{\Delta^2+\Gamma^2}\left(\dfrac13\rho^{22}_{11}+\rho^{11}_{11}\right),
\end{equation}
\end{subequations}

\noindent where upper indices ``u,'' ``d'' of the density matrix elements denote upper-state levels $F_e=2,1$ and indices ``2,'' ``1'' denote ground-state levels $F_g=2,1$. Lower indices denote $m_F$ value.

Equations for elements of the ground state are the following:
\onecolumngrid
\begin{subequations}
\begin{equation}
\Gamma_g\rho^{22}_{-2-2}=-\Gamma\left[\dfrac13\dfrac{V^2}{\Delta^2+\Gamma^2}+\dfrac{V^2}{(\Delta+\omega_e)^2+\Gamma^2}\right]\rho^{22}_{-2-2}+\dfrac{\Gamma_g}{8}+\gamma\left(\dfrac13\rho^{uu}_{-2-2}+\dfrac16\rho^{uu}_{-1-1}+\dfrac12\rho^{dd}_{-1-1}\right),
\end{equation}
\begin{equation}
\Gamma_g\rho^{22}_{-1-1}=-\dfrac12\Gamma\left[\dfrac{V^2}{\Delta^2+\Gamma^2}+\Gamma\dfrac{V^2}{(\Delta+\omega_e)^2+\Gamma^2}\right]\rho^{22}_{-1-1}\\
+\dfrac{\Gamma_g}{8}+\gamma\left(\dfrac16\rho^{uu}_{-2-2}+\dfrac{1}{12}\rho^{uu}_{-1-1}+\dfrac14\rho^{uu}_{00}+\dfrac14\rho^{dd}_{-1-1}+\dfrac14\rho^{dd}_{00}\right),
\end{equation}
\begin{equation}
\Gamma_g\rho^{11}_{-1-1}=-\dfrac16\Gamma\left[\dfrac{V^2}{\Delta^2+\Gamma^2}+\Gamma\dfrac{V^2}{(\Delta+\omega_e)^2+\Gamma^2}\right]\rho^{11}_{-1-1}+\dfrac{\Gamma_g}{8}+\gamma\left(\dfrac12\rho^{uu}_{-2-2}+\dfrac14\rho^{uu}_{-1-1}+\dfrac{1}{12}\rho^{uu}_{00}+\dfrac{1}{12}\rho^{dd}_{-1-1}+\dfrac{1}{12}\rho^{dd}_{00}\right),
\end{equation}
\begin{equation}
\begin{gathered}
\Gamma_g\rho^{22}_{00}=-\dfrac12\Gamma\left[\dfrac{V^2}{\Delta^2+\Gamma^2}+\dfrac13\Gamma\dfrac{V^2}{(\Delta+\omega_e)^2+\Gamma^2}\right]\rho^{22}_{00}\\
+\dfrac{1}{2}\dfrac{V^2}{\Delta^2+\Gamma^2}\left[\Delta\cdot\text{Im}(\rho^{21}_{00})+\Gamma\cdot\text{Re}(\rho^{21}_{00})\right]+\dfrac{1}{6}\dfrac{V^2}{(\Delta+\omega_e)^2+\Gamma^2}\left[\left(\Delta+\omega_e\right)\cdot\text{Im}(\rho^{21}_{00})+\Gamma\cdot\text{Re}(\rho^{21}_{00})\right]\\
+\dfrac{\Gamma_g}{8}+\gamma\left(\dfrac14\rho^{uu}_{-1-1}+\dfrac14\rho^{uu}_{11}+\dfrac{1}{12}\rho^{dd}_{-1-1}+\dfrac13\rho^{dd}_{00}+\dfrac{1}{12}\rho^{dd}_{11}\right),
\end{gathered}
\end{equation}
\begin{equation}
\begin{gathered}
\Gamma_g\rho^{11}_{00}=-\dfrac12\Gamma\left[\dfrac{V^2}{\Delta^2+\Gamma^2}+\dfrac13\dfrac{V^2}{(\Delta+\omega_e)^2+\Gamma^2}\right]\rho^{11}_{00}\\
-\dfrac{1}{2}\dfrac{V^2}{\Delta^2+\Gamma^2}\left[\Delta\cdot\text{Im}(\rho^{21}_{00})-\Gamma\cdot\text{Re}(\rho^{21}_{00})\right]-\dfrac{1}{6}\dfrac{V^2}{(\Delta+\omega_e)^2+\Gamma^2}\left[\left(\Delta+\omega_e\right)\cdot\text{Im}(\rho^{21}_{00})-\Gamma\cdot\text{Re}(\rho^{21}_{00})\right]\\
+\dfrac{\Gamma_g}{8}+\gamma\left(\dfrac14\rho^{uu}_{-1-1}+\dfrac13\rho^{uu}_{00}+\dfrac14\rho^{uu}_{11}+\dfrac{1}{12}\rho^{dd}_{-1-1}+\dfrac{1}{12}\rho^{dd}_{11}\right),
\end{gathered}
\end{equation}
\begin{equation}
\Gamma_g\rho^{22}_{11}=-\dfrac13\Gamma\dfrac{V^2}{\Delta^2+\Gamma^2}\rho^{22}_{11}+\dfrac{\Gamma_g}{8}+\gamma\left(\dfrac16\rho^{uu}_{22}+\dfrac{1}{12}\rho^{uu}_{11}+\dfrac14\rho^{uu}_{00}+\dfrac14\rho^{dd}_{11}+\dfrac14\rho^{dd}_{00}\right),
\end{equation}
\begin{equation}
\Gamma_g\rho^{11}_{11}=-\Gamma\dfrac{V^2}{\Delta^2+\Gamma^2}\rho^{11}_{11}+\dfrac{\Gamma_g}{8}+\gamma\left(\dfrac12\rho^{uu}_{22}+\dfrac14\rho^{uu}_{11}+\dfrac{1}{12}\rho^{uu}_{00}+\dfrac{1}{12}\rho^{dd}_{11}+\dfrac{1}{12}\rho^{dd}_{00}\right),
\end{equation}
\begin{equation}
\Gamma_g\rho^{22}_{22}=\dfrac{\Gamma_g}{8}+\gamma\left(\dfrac13\rho^{uu}_{22}+\dfrac16\rho^{uu}_{11}+\dfrac12\rho^{dd}_{11}\right),
\end{equation}
\begin{equation}
\begin{gathered}
\left\{\delta+i\Gamma_g+\dfrac{i}{2}\Gamma\left[\dfrac{V^2}{\Delta^2+\Gamma^2}+\dfrac{1}{3}\dfrac{V^2}{(\Delta+\omega_e)^2+\Gamma^2}\right]\right\}\rho^{21}_{00}=\\
\dfrac{i}{4}\Gamma\left[\dfrac{V^2}{\Delta^2+\Gamma^2}+\dfrac{1}{3}\dfrac{V^2}{(\Delta+\omega_e)^2+\Gamma^2}\right](\rho^{22}_{00}+\rho^{11}_{00})+\dfrac14\left[\Delta\dfrac{V^2}{\Delta^2+\Gamma^2}+\dfrac{1}{3}(\Delta+\omega_e)\dfrac{V^2}{(\Delta+\omega_e)^2+\Gamma^2}\right](\rho^{22}_{00}-\rho^{11}_{00}).
\end{gathered}
\label{coherence}
\end{equation}
\label{groundstate}
\end{subequations}
\twocolumngrid

The light shift of the ground-state microwave transition frequency is neglected in system of equations~\eqref{groundstate} to simplify calculations. To account for complete depolarization of the excited state, we replaced all its populations with the arithmetic mean: $\rho^{uu}_{ii},\,\rho^{dd}_{ii}\rightarrow\left(\sum^{2}_{i=-2}\rho^{uu}_{ii}+\sum^{1}_{i=-1}\rho^{dd}_{ii}\right)/8$. Solution for significant rate of optical pumping, $V^2/\Gamma\gg\Gamma_g$, demonstrated that physical contrast, which we define here as $\left[\rho^{ee}(|\delta|\gg V^2/\Gamma)-\rho^{ee}(\delta=0)\right]/\rho^{ee}(|\delta|\gg V^2/\Gamma)$, is two-times greater when the excited state is depolarized. Here $\rho^{ee}$ is the sum of populations of the excited-state sublevels.

Fig.~\ref{LevelsScheme} demonstrates distributions of populations of $^{87}$Rb $5$S$_{1/2}$ and $5$P$_{1/2}$ states for two cases: without (a) and with (b) excited-state depolarization. They were calculated for $\Gamma/2\pi=1$~GHz, $\omega_e/2\pi=817$~MHz, $\Delta/2\pi=-30$~MHz, $\delta=0$. The value of Rabi frequency was set to provide power broadening of the CPT resonance three-times greater than $\Gamma_g$. In case (b) population of sublevel \hbox{$m_{F_e}=2$} decreases and optical pumping of the non-absorbing sublevel \hbox{$m_{F_g}=2$} becomes smaller. Populations of excited-state sublevels \hbox{$F_e=2,\,m_{F_e}=-2,-1,0$}, \hbox{$F_e=1,\,m_{F_e}=-1,0$} grow, which increases amount of spontaneous transitions to working sublevels \hbox{$F_g=1,2,\,m_{F_g}=0$}. We note that population of sublevel \hbox{$F_g=1,\,m_{F_g}=1$} is smaller than that of \hbox{$F_g=1,\,m_{F_g}=-1$}, since probability of transition \hbox{$|F_g=1,\,m_{F_g}=1\rangle\rightarrow|F_e=2,\,m_{F_e}=2\rangle$ is~greater}, while the repopulation rate of these sublevels is the same due to spontaneous transitions.

We also note that equation~\eqref{coherence} for coherence $\rho^{21}_{00}$ contains terms $\propto(\rho^{22}_{00}-\rho^{11}_{00})$ in its right-hand side. Despite that components of bichromatic field have equal intensities, populations of working sublevels are not equal due to spontaneous transitions. As a consequence, the real part of coherence $\rho^{21}_{00}$ acquires a term proportional to $\delta$. The CPT resonance becomes neither an even nor an odd function of $\delta$, i.e., it turns out to be asymmetric. On the opposite, under complete depolarization of the excited state, spontaneous transitions equally populate working sublevels providing a symmetric CPT resonance.

\bibliography{references}

\end{document}